\documentclass[twocolumn,showpacs,floatfix,amsmath,amssymb,prb]{revtex4}
\usepackage[dvips]{graphicx}
\usepackage{latexsym}
\usepackage{graphicx}
\usepackage{times,psfrag,subfigure}
\usepackage{amsmath}
\usepackage{dcolumn}
\newcommand{\la}{\langle}
\newcommand{\ra}{\rangle}
\newcommand{\nn}{{\nonumber}}
\newcommand{\br}{{\bf r}}
\newcommand{\bk}{{\bf k}}
\newcommand{\bK}{{\bf K}}
\newcommand{\bj}{{\bf j}}
\newcommand{\bE}{{\bf E}}
\newcommand{\hz}{\hat{z}}
\newcommand{\be}{\begin{equation}}
\newcommand{\ee}{\end{equation}}
\newcommand{\bea}{\begin{eqnarray}}
\newcommand{\eea}{\end{eqnarray}}

\begin{document}

\title{Transition to zero resistance in a two dimensional electron
gas driven with microwaves}
\author{Jason Alicea}
\email{aliceaj@physics.ucsb.edu}
\affiliation{Physics Department, University of California, Santa
Barbara, CA 93106}
\author{Leon Balents}
\email{balents@physics.ucsb.edu}
\affiliation{Physics Department, University of California, Santa
Barbara, CA 93106}
\author{Matthew P. A. Fisher}
\email{mpaf@kitp.ucsb.edu}
\affiliation{Kavli Institute for Theoretical Physics, University of
California, Santa Barbara, CA 93106}
\author{Arun Paramekanti}
\email{arun@kitp.ucsb.edu}
\affiliation{Kavli Institute for Theoretical Physics, University of
California, Santa
Barbara, CA 93106}
\author{Leo Radzihovsky}
\email{radzihov@colorado.edu}
\affiliation{Department of Physics, University of Colorado, Boulder,
CO 80309}

\date{\today}

\begin{abstract}
High-mobility 2D electron systems in a perpendicular 
magnetic field exhibit zero resistance states (ZRS) when driven with 
microwave radiation. We study the nonequilibrium phase transition into 
this ZRS using phenomenological equations of motion
to describe the current and density fluctuations. We focus on
two models for the transition into a time-independent steady state.
Model-I assumes rotational invariance, density
conservation, and symmetry under shifting the density globally by a
constant. This model is argued to describe physics on small length
scales where the density does not vary appreciably from its mean.
The ordered state that arises in this case breaks rotational
invariance and consists of a uniform current and transverse Hall
field.  We discuss some properties of this state, such as
stability to fluctuations and the appearance of a
Goldstone mode associated with the continuous symmetry breaking.
Using dynamical renormalization group techniques, we find
that with short-range interactions this model can admit a continuous
transition described by mean-field theory, whereas with long-range
interactions the transition is driven first-order.   Model-II,
which assumes only rotational invariance and density
conservation and is argued to be appropriate on longer length scales,
is shown to predict a first-order transition with either short-
or long-range interactions.  We discuss implications for experiments,
including scaling relations and a possible way to detect
the Goldstone mode in the case of a continuous transition into the ZRS, 
as well as possible signatures of a first-order transition in larger samples.
We also point out the connection of our work to the well-studied
phenomenon of `flocking'.
\end{abstract}

\maketitle

\section{introduction}

High-mobility two-dimensional electron gases (2DEGs) subject to a
perpendicular magnetic field exhibit novel physics when driven with
microwave radiation.  Zudov \emph{et al}.\cite{Zudov1} first
demonstrated that the longitudinal resistance develops dramatic
radiation-induced oscillations at low temperatures ($T\sim 1K$) and
low magnetic fields ($B \lesssim 1kG$).  These oscillations are
periodic in $1/B$, with the period set by the ratio of the microwave
and cyclotron frequencies. In contrast, the Hall resistance is nearly
unaffected by the microwaves\cite{Zudov1}, although small
radiation-induced Hall oscillations have recently been
observed\cite{HallOsc,HallOsc2}.  More spectacular is the observation,
made independently by Mani \emph{et al}.\cite{Mani} and Zudov \emph{et
al}.\cite{Zudov2}, that in even higher-mobility samples the
oscillations become sufficiently large that the minima of the
resistance oscillations develop into zero resistance states --- the
measured resistance vanishes within experimental accuracy over a range
of magnetic fields and radiation intensities. Subsequent experiments
have confirmed their results\cite{HallOsc,Dorozhkin, Willett} and also
observed a similar effect in Corbino samples\cite{Corbino}, where
zero-conductance states have been measured.

On the theoretical front, several groups have carried out microscopic
calculations of the resistance taking into account photon-assisted impurity
scattering\cite{Ryzhii1,Ryzhii2,Durst,Lei1,Lei2,Vavilov} and
radiation-induced changes in the distribution
function\cite{Dmitriev1,Dmitriev2}.
Although these mechanisms are quite different, both capture the
resistance oscillations with the correct period and phase at low radiation
intensity. At higher intensities, however, these calculations predict a
\emph{negative} resistance in regions of magnetic field where the
experiments find a zero resistance state.

The missing ingredient needed to connect the microscopic theory with
the experiments was the observation of Andreev \emph{et al}.\ that a
state characterized by a negative longitudinal resistance, quite
independent of its microscopic origin, is unstable to current
fluctuations.\cite{Andreev} They argued that this instability leads to
an inhomogeneous state where the system spontaneously develops domains
of current with magnitude $j_0$, where $j_0$ corresponds to a
vanishing longitudinal resistivity, i.e. $\rho_D(j_0^2) = 0$.
Applying an external current then merely reorganizes the domain sizes
in order to accommodate the additional current, leading to zero
measured resistance over a range of bias current as observed
experimentally. In the absence of an applied current, since the system
has a large Hall resistance, this picture indicates that spontaneous
current domains in the zero resistance state should reveal themselves
through spontaneous Hall voltages transverse to the domains. Willett
\emph{et al}.\cite{Willett} have indeed measured spontaneous voltages
between internal and external contacts with no applied current, which
lends support to this idea.

The experiments together with the microscopic calculations and
phenomenological arguments provide strong evidence for the existence
of a
\emph{nonequilibrium} phase transition from a normal state with nonzero
resistance to a zero resistance state whose detailed properties remain
largely unexplored.  In this paper, we attempt to gain an understanding of
the nature of this transition, and to learn about the properties of the zero
resistance state.

\subsection{Strategy}

We begin with the observation that while the microscopic mechanism for how
radiation induces the transition to a zero resistance state is a matter of
some debate, this knowledge is not crucial for studying universal properties
close to the phase transition. Indeed, in order to study the long-wavelength,
low-frequency dynamics near the transition it is sufficient to identify the
appropriate hydrodynamic variables and construct the most general local
equations of motion for them consistent with symmetries and conservation
laws. The magnetic field, temperature, microwave radiation, and quantum
effects will determine the various parameters of this theory; these may be
calculated in principle from a microscopic approach, but we do not
attempt to do this
here. Our idea will be to view the equations of motion as a non-equilibrium
analogue of Landau-Ginzburg-Wilson theory.  We will use them
to study universal physics near the phase transition, 
going beyond mean field
theory by including nonlinearities and fluctuations within a renormalization
group framework.

In the vicinity of the transition into the zero resistance state in
the 2DEG, the relevant hydrodynamic degrees of freedom are the
current density $\bj(\br,t)$ and the charge density $n(\br,t)$, which are
constrained by a continuity equation,
\begin{equation}
  \frac{\partial n}{\partial t} + \nabla\cdot\bj = 0,
  \label{continuity}
\end{equation}
that enforces local charge conservation.  

The dynamics of the current density $\bj(\br,t)$ is governed by a
nonequilibrium equation of motion (akin to the Navier-Stokes equation)
for a 2D charged fluid in a perpendicular magnetic field.  Because of
the nonequilibrium nature of the system (microwave-driven 2D electron
liquid) the equation for $\bj$ includes non-conservative forces, i.e.,
those not derivable from a free energy functional.  Hence, the generic
symmetry-allowed form of the equation for the current density is only
restricted by the translational and rotational invariances in the
plane. Keeping the leading order (at long length and time scales)
terms in powers of the charge and current densities and their
gradients leads to
\bea
  \omega_0^{-1}\partial_t^2 { \bf j} &+& \partial_t {\bf j} =
  -r{\bf j} - u|{\bf j}|^2{\bf j} + \eta_1 \nabla^2 {\bf j} + \eta_2
  \nabla(\nabla\cdot {\bf j})
  \nonumber \\
  &-& \eta_3 \nabla^4 {\bf j} - \eta_4 \nabla^3(\nabla \cdot {\bf j})
  + \tilde \omega_c {\bf \hat z} \times {\bf j}
  - \mu\nabla \Phi
  \nonumber \\
  &-& \nu_1({\bf j} \cdot \nabla){\bf j}
  -\nu_2\nabla{\bf j}^2
  - \nu_3 (\nabla \cdot {\bf j}){\bf j} + \gamma_1
  \Phi\nabla\Phi
  \nonumber \\
  &+& \gamma_2 \Phi {\bf j}
  + \gamma_3 \Phi {\bf \hat z} \times {\bf j}
  + {\bf \zeta} + \cdots.
\label{jeq}
\eea
As we will discuss in more detail below, terms appearing on the
right-hand side of the above equation are forces that determine the
local acceleration ($\partial_t\bj(\br,t)$) of the electron fluid,
each having a simple physical interpretation. The $r$ and $u$ terms
are the linear and nonlinear longitudinal resistivities (frictional
drag forces on the electron fluid). The $\eta_i$ terms describe viscous
forces associated with a nonuniform flow and the $\tilde{\omega}_c$
term is the Lorentz force on the charged moving electron fluid. The
$\nu_i$ terms are convective-like nonlinearities, where the absence of
Galilean invariance permits more general types of convective terms
$\nu_2$ and $\nu_3$ in addition to the conventional $\nu_1$ term,
with generic (symmetry unrestricted) values of these couplings.

Here, the potential $\Phi \equiv \Phi[n]$ is determined by the
density via
\begin{equation}
  \Phi = \int_{\br'} V(\br-\br') n(\br').
  \label{potential}
\end{equation}
For long-range interactions, $V(\br-\br') \sim 1/|\br-\br'|$ is the
Coulomb potential. For a screened interaction, we can set $V(\br-\br')
\approx \delta(\br-\br')$, so that $\Phi \approx n$. With this, the $\mu$ 
and $\gamma_1$ terms incorporate Fick's law (diffusion down a local
chemical potential gradient), with the latter accounting for a
density-dependent diffusion coefficient. Similarly, $\gamma_2$ and
$\gamma_3$ account for the lowest order density-dependence of the
linear resistivity and the Lorentz force.

In addition, we have included in Eq.\ (\ref{jeq}) a zero-mean white noise
force ${\bf \zeta}$ with a correlator
\begin{equation}
\la\zeta^\alpha(\br,t) \zeta^\beta(\br',t') \ra = 2 g
\delta^{\alpha\beta} \delta(\br-\br') \delta(t-t').
\label{noise}
\end{equation}
Apart from thermal noise, this incorporates the effect
of microscopic fluctuations that arise from the coarse graining implicit
in our formulation. Since we are dealing with a system far from
equilibrium, the strength $g$ of the noise is not fixed by the
fluctuation-dissipation relation, but is an independent quantity.

Focusing on the terms $\partial_t {\bf j} = - r{\bf j} + {\bf \zeta}$
in Eq.~(\ref{jeq}), it is clear that (i) for large positive values of
$r$, the zero current state is stable and current fluctuations decay
exponentially, while (ii) for large negative values of $r$, current
fluctuations grow exponentially and the zero current state is
unstable.  Thus, as $r$ changes from positive to negative (in the
experiments tunable by a microwave power and/or frequency),
Eqs.~(\ref{continuity}-\ref{noise}) describe the phase transition from
a conventional resistive state for $r>0$ to a nonequilibrium steady
state with spontaneous currents for $r<0$.

As we will argue in Section II, this set of equations can potentially
describe various types of current and density ordering, including
circulating current states and domain patterns of current and
density. In this paper our main focus will be on the nature of the
transition into a time-independent steady state with possible density
and current domains since, given the observations of Willett \emph{et
al.}, this appears to be relevant to the 2DEG experiments.  We defer to
future work questions regarding the detailed nature of the ordered state
in this case, as well as a study of the phase transition into the
circulating state.

A quite different theoretical motivation for studying this problem
arises from the observation that the current and density evolution
equations studied here reduce, for $B=0$ and short range interactions,
to the continuum equations used to investigate the problem of
`flocking'.\cite{Toner1,Toner2} In that case, the system has been
shown to develop an expectation value for the particle current, thus
spontaneously breaking the continuous rotational symmetry even in two
spatial dimensions. This is particularly striking since the
Mermin-Wagner theorem\cite{MerminWagner} forbids such symmetry
breaking in $d=2$ for classical equilibrium systems. This `violation'
was identified as arising from nonlinear convective terms which are
only allowed in non-equilibrium systems, and turn out to be relevant
for this problem in dimensions $d < 4$. Much is known about the
universal dynamics in the flocking state in $d=2$, but the nature of
the phase transition into this state has not been addressed
analytically.  The question we study is equivalent to asking: What is
the fate of the flocking transition and the flocking state in two
dimensions in the presence of a magnetic field that breaks
time-reversal symmetry? As we show, one can make more progress in this
modified problem. This `flocking' point of view is also useful for
carrying out numerical simulations, since many simple particle models
for flocking have been studied in the absence of a magnetic field and
can be adapted to our problem, although we do not pursue this here.

\subsection{Summary of the paper}

We begin in Section II by showing how some terms in Eq.~(\ref{jeq}) can be
related to the full nonlinear resistivity. We do this by formally expanding
the relation
\begin{equation}
  \bE(\bk,\omega) = \rho_D(\bj,\Phi) \bj + \rho_H(\bj,\Phi)\bj \times \hz
  \label{cartooneqn}
\end{equation}
at low frequency and wavevector, and for small current and potential
fluctuations.  Here $\rho_D$ and $\rho_H$ represent the diagonal and
Hall resistivities, and the electric field $\bE$ is determined via the
electrostatic potential, i.e.  $\bE = -\nabla \Phi$.  Upon Fourier
transforming back to real space, one can arrive at an equation with a
form similar to Eq.\ (\ref{jeq}).  This proves to be a useful exercise
since we can then relate different possible forms of the frequency and
wavevector dependent resistivity in the presence of microwaves to the
model parameters appearing in Eq.\ (\ref{jeq}) and therefore to the kinds
of ordered states which might emerge from our description. More
importantly, this helps us to identify the correct set of critical
modes near the phase transition into these putative ordered
states. Specifically, we show that if the resistivity is an increasing
function of frequency at low frequency, so that the zero resistance
state is achieved when the DC resistance first goes negative, then a
time-independent steady state with inhomogeneous density would
result. The only critical mode near the transition into this state
involves density fluctuations accompanied by current fluctuations that
balance the Lorentz force.  Since the current and density are tied to
one another in this mode, one can re-express current fluctuations in
terms of the density.  Inserting the resulting expression into the
continuity equation results in an equation of motion involving only
the density at the critical point.

This equation of motion for the density at the critical point depends
on terms involving the absolute magnitude of the density, as well as
terms that depend only on density gradients. We warm up in Section III
by analyzing a model which neglects terms that depend on the absolute
magnitude of the density, and is instead invariant under shifting the
density by a constant.  This model is expected to describe physics on
short length scales where the density does not vary appreciably from
its mean so that such terms can be safely ignored.  In the `ordered
phase' of this model the system develops a uniform current with a
transverse density gradient that balances the Lorentz force.  We show
that this state is stable to small fluctuations, and discuss the
`Goldstone mode' associated with the spontaneously broken rotational
symmetry.  The ordered state described by this model is argued to be
relevant for the experiments at short length scales $L < L_{\rm c1}$,
where $L_{\rm c1}$ is estimated to be roughly 1mm, comparable to
sample sizes used in the experiments.  We then turn to the critical
properties of this model, considering both short- and long-range
interactions.  With short-range interactions, we show that the upper
critical dimension is $d_{uc} = 2$.  In this case, we use dynamical
renormalization group calculations to demonstrate that the Gaussian fixed
point has a finite-volume basin of attraction; hence, a finite
fraction of initial nonlinear couplings all flow to zero upon
renormalization.  In such cases, the transition is \emph{continuous}
and governed to a good approximation by mean field theory.  Various
scaling relations should hold near the transition in this regime.  For
instance, at fixed magnetic field strength and in the absence of an
applied voltage, below the transition, the spontaneous current $j_0$
should scale with the microwave power $P$ as
\begin{equation}
  j_0(P) \sim |P-P_c|^\beta,
\end{equation}
where $\beta>0$ and $P_c$ is the critical microwave power at which the
longitudinal resistance first vanishes. Approaching the transition
from the resistive side, with $P<P_c$, we expect a universal scaling
relation to hold between the imposed current $j$ and the induced
longitudinal electric field,
\begin{equation}
  j(P,E) \sim (P_c-P)^\beta f\left(E^{1/\delta} (P_c-P)^{-\beta}\right),
\end{equation}
where $f(x)$ is a scaling function with the properties that $f(x)
\sim x^\delta$ as $x\to 0$,
and $f(x) \sim x$ as $x \to \infty$. The behavior as $x\to 0$ recovers
the linear response result, $j \propto E$, in the resistive phase, with a
resistivity $\sim (P_c-P)^{\beta(1-\delta)}$.
The behavior for $x\to \infty$
leads to a universal longitudinal nonlinear IV characteristic
\begin{equation}
  j(P_c,E) \sim E^{1/\delta}
\end{equation}
at the transition, $P=P_c$, with mean-field value of $\delta = 3$ for
a current-biased experiment.  We then show that long-range
interactions appear to drive the transition first-order. The
experimental signatures of the Goldstone mode in the `ordered phase'
and the mean field transition with short range interactions are
qualitatively discussed in Section V.

In Section IV we analyze the phase transition in the more general
model, where terms that depend on the absolute magnitude of the
density are taken into account. These terms, which become important
on length scales $L > L_{\rm c2}$, are argued to drive the
transition first-order with either short- or long-range interactions
based on renormalization group calculations. We derive an expression
for $L_{\rm c2}$ that depends on the density- and
wavevector-dependent resistivity, and suggest that microscopic
calculations may be used to estimate this length. Experimental
consequences of the first-order phase transition for sample sizes
larger than $L_{\rm c2}$ are briefly noted in Section V.

\section{Deriving and simplifying the equations of motion}

\subsection{``Microscopic derivation'' of equations of motion}

Before we turn to the analysis of the phases and transitions described
by the set of Eqs.\ (\ref{continuity}-\ref{noise}), let us consider a
derivation of some terms in the equation of motion for $\bj$ in
Eq.\ (\ref{jeq}).

We begin with the linear response relation
\begin{equation}
  \bE^\alpha(\bk,\omega) =
  \rho_D^{\alpha\beta}(\bk,\omega) \bj^\beta(\bk,\omega) + \rho_H(\bk,\omega)
  \epsilon^{\alpha\beta} \bj^\beta(\bk,\omega)
  \label{linearresponse}
\end{equation}
where $\rho_D$ and $\rho_H$ represent the diagonal and Hall
resistivities, the electric field $\bE$ is determined from the
electrostatic potential via $\bE = -\nabla \Phi$, and
$\epsilon^{\alpha\beta}$ is the antisymmetric tensor.

We know from experiments that $\rho_H(0,0) \approx B/(ne)$ even
in the presence of microwave radiation.
We are interested in the case where the dissipative
part of the microscopic diagonal resistivity becomes negative.
With increasing microwave intensity, this would first happen at
some particular wavevector and frequency $(\bK,\Omega)$. Two
specific cases for the behavior of $\rho_D(\bk,\omega)$ are
illustrated in Fig.~\ref{rhofig}.

If $\bK,\Omega$ are small, we can access the resistivity
minimum shown in Fig.~\ref{rhofig} by expanding
$\rho_D(\bk,\omega) = \rho_1(\bk,\omega) + i\rho_2 (\bk,\omega)$
in a Taylor series as:
\bea
    \rho_D^{\alpha\beta}&=& \delta^{\alpha\beta} \left[ \rho_1(0,0)
    + i \omega (\frac{\partial\rho_2}{\partial \omega}) +
    \frac{\omega^2}{2}
    (\frac{\partial^2\rho_1}{\partial\omega^2}) + \bar{\eta}_1 \bk^2\right]
    \nn\\
    &+& \bar{\eta}_2 \bk^\alpha \bk^\beta + \cdots,
\eea
where the frequency derivatives and coefficients $\bar
\eta_{1,2}$ are evaluated at $(\bk=0,\omega=0)$. Using this expansion
inside Eq.\ (\ref{linearresponse}), and assuming that the Hall
resistivity is independent of wavevector and frequency in the regime
of interest, we find the following relations between the coefficients
in Eq.~(\ref{jeq}) and the microscopic linear response resistivity,
\bea
    r &=& \rho_1(0,0)/G \\
    \omega_0^{-1}&=&\frac{1}{2 G} \frac{\partial^2\rho_1}{\partial\omega^2} \\
    \mu &=& 1/G \\
    \eta_1 &=& \bar{\eta}_1/G \\
    \eta_2 &=& \bar{\eta}_2/G \\
    \tilde{\omega}_c&=&\rho_H(0,0)/G,
\eea
where $G \equiv -(\partial\rho_2/\partial\omega)$.

We can similarly match some of the non-linear terms in
Eq.~(\ref{jeq}) as follows. Let us take $\bk=0,\omega=0$ and
consider the nonlinear resistivity that depends in general on the
local potential and the current magnitude, namely,
\bea
    \rho_D^{\alpha\beta}(\bj^2,\Phi) &=& \delta^{\alpha\beta}
    \left[\rho_1(0,0) + \bar{u} \bj^2 - \bar{\gamma_2} \Phi + \cdots\right]\\
    \rho_H(\bj^2,\Phi) &=&
    \rho_H(0,0) + \bar{u}_H \bj^2 - \bar{\gamma_3} \Phi + \cdots.
\eea
Using this expansion and comparing with the non-linear terms in
Eq.~(\ref{jeq}), we find
\bea
    u &=& \bar{u}/G \\
    \gamma_2&=&\bar{\gamma}_2/G \\
    \gamma_3&=&\bar{\gamma}_3/G.
\eea As we shall see below, the type of ordering expected to emerge
from our description depends on the frequency and wavevector
dependence of the resistivity -- measuring these in the disordered
phase close to the transition would offer clues to the nature of the
zero resistance state.

\subsection{Identifying critical modes and simplifying the
equations of motion}

\begin{figure}
  \begin{center}
    {\resizebox{8cm}{!}{\includegraphics{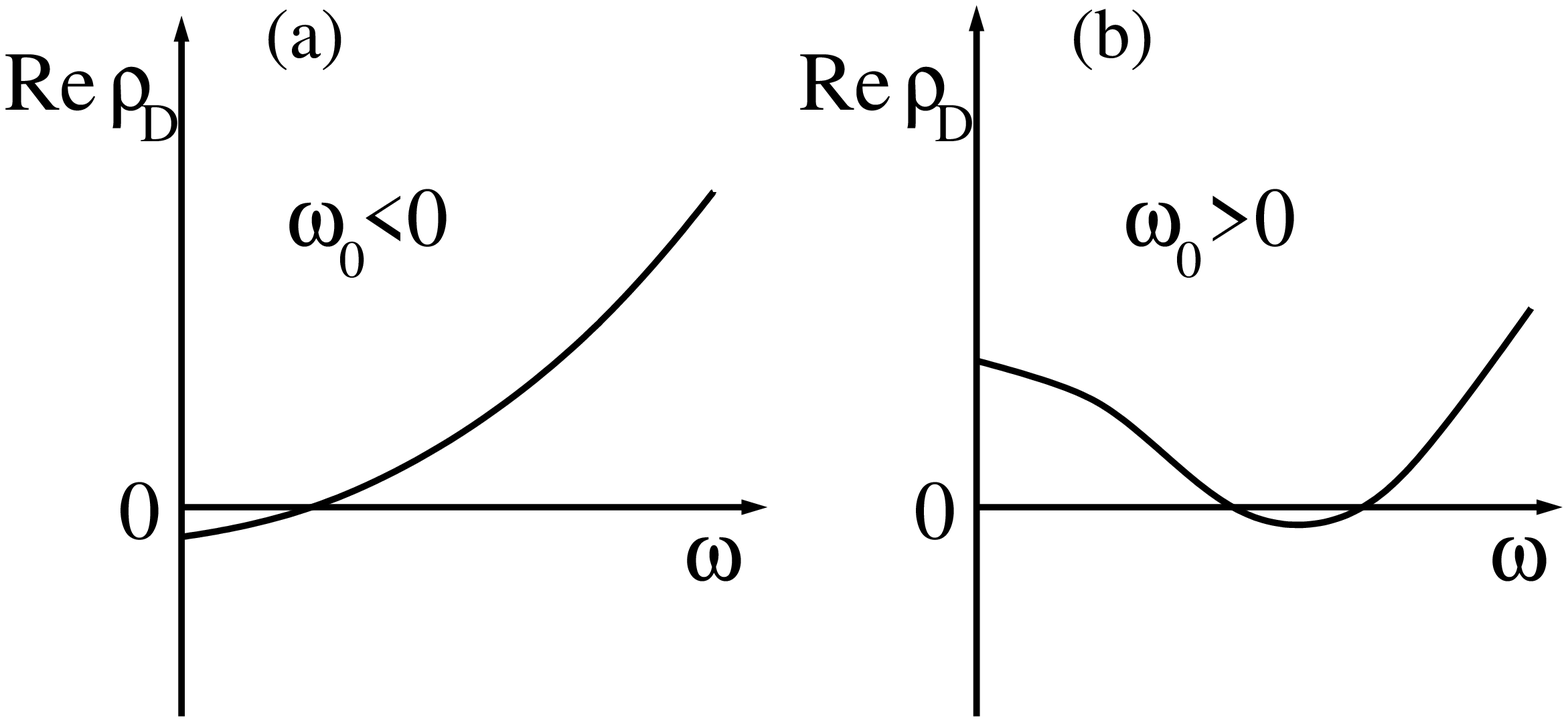}}}
  \end{center}
  \caption{Schematic behavior of the real part of $\rho_D$
when (a)
    $\omega_0 <0$ and (b) $\omega_0 >0$.}
  \label{rhofig}
\end{figure}

On general grounds, one would expect that the type of order that
develops near the transition should depend on where the
minimum of $\rho_1$ occurs in ($k,\omega$) space.
The resistance will in general depend on both $k$ and $\omega$, and
close to the transition will only be negative in a
small region of frequencies and wavevectors about the minimum.  Modes
away from the minimum remain stable.  Two
possibilities for where this minimum occurs as a function of frequency
are sketched in Fig.~\ref{rhofig}.  If the minimum occurs at zero
frequency as in Fig.~\ref{rhofig}(a), then zero-frequency modes that
become critical at the transition should give rise to a time-independent
ordered state (e.g., static domains of current).  If on the other hand
the minimum occurs at a non-zero frequency as in Fig.~\ref{rhofig}(b),
then finite-frequency modes should give rise to a state ordered at
finite frequency (e.g., circulating currents).
In either case, the wavevector at
which the minimum occurs would determine the wavevector at which the
system orders.  Thus, the signs of $\eta_1$ and $\omega_0$, which
determine whether the minimum of
$\rho_1$ occurs at zero (or nonzero) wavevector and frequency,
should play an important role in the ordering.

To make this more concrete, let us consider the mode structure
in the disordered state, where ${\bf j}$ and $n$
represent fluctuations about a stable zero-current state.
We start with the case $\eta_1 >0$ and focus on wavevectors
$k\rightarrow 0$ since the resistivity is minimized when $k = 0$.
The modes obtained from the linearized equations of motion are given by
\begin{eqnarray}
  \omega_\pm &=& -i\bigg{(}r+\frac{r^2-\omega_c^2}{\omega_0}\bigg{)}
     \pm\omega_c\bigg{(}1+\frac{2r}{\omega_0}\bigg{)}
  \nonumber \\
  &+& O(k^2V (k),\omega_0^{-2})
  \label{Bmodes}
  \\
  \omega_D &=& \frac{-i r}{r^2+\omega_c^2}\mu V(k)k^2 +O(k^4 V^2(k)),
  \label{nmode}
\end{eqnarray}
where $V(k)$ is the Fourier-transform of the interaction potential
$V(r)$.  Equation (\ref{Bmodes}) is only written out to order
$\omega_0^{-1}$ for simplicity.  We only want to consider here the
effect of adding a small frequency dependence to the resistivity, so
the exact expression is not important.  The modes in Eq.\
(\ref{Bmodes}) correspond to current fluctuations that circulate due
to the magnetic field as they dissipate.  The associated density
fluctuations for these modes vanish in the $k\rightarrow 0$ limit.
Equation (\ref{nmode}) represents a diffusive mode involving both
current and density fluctuations that survive in the $k\rightarrow 0$
limit.  These current fluctuations are undeflected by the magnetic
field because the Lorentz force is balanced by an electric field set
up by density fluctuations.

In order for the zero-current state to be stable, the imaginary part
of these frequencies must be negative so that fluctuations are
damped exponentially in time.  For the diffusive mode, stability
requires $r\geq 0$.  The circulating current modes are stable when
$r\gtrsim\omega_c^2/\omega_0$, assuming $\omega_0^2 \gg \omega_c^2$
for simplicity.  Violation of either inequality renders the
zero-current state of the system unstable to current fluctuations.
Since $r \sim \rho_1$, this instability occurs approximately where
the longitudinal resistivity changes sign, consistent with the
findings of Andreev \emph{et al}.\cite{Andreev}

As the longitudinal resistance tends to zero and the ordered state is
approached, the
circulating current modes become critical before the diffusive mode
if $\omega_0>0$.  (Note that since these modes propagate at a finite
frequency, this is consistent with the above discussion on the
frequency-dependence of the longitudinal resistivity.)
Once the circulating current modes become unstable,
the system should undergo a transition
into an ordered state where circulating currents spontaneously
develop but the density
remains uniform.  If $\omega_0<0$, however, the diffusive mode
becomes critical while the circulating current modes remain damped.
In this case one would expect the system to undergo
a transition into a phase with nonuniform density
and spontaneous currents ordered at zero wavevector.
A distinguishing characteristic of the
latter phase would be the development of voltages resulting from the
nonuniform density.  Since spontaneous
voltages in the absence of a net current have indeed been
observed in the ordered state, the case $\omega_0<0$
seems to be the experimentally relevant one.  We consequently focus on
the transition into the density-ordered state and leave an analysis
of the circulating-current state to future studies.

These same ideas can be applied to the case $\eta_1 < 0$, where the
resistivity is minimized at finite wavevector.  Assuming $\eta_3
>0$, one is then interested in wavevectors with magnitude close to
$k_0 = (|\eta_1|/2\eta_3)^{1/2}$, corresponding to the resistivity
minimum. Since we can no longer perturb in $k$, we cannot in
general write down simple expressions for the modes in the
disordered state.  We will therefore focus on the point where the
resistance at zero frequency and $k = k_0$ drops to zero since this
simplifies the mode structure.  (This happens when $r =
\eta_1^2/4\eta_3$.) A critical diffusive mode then emerges whose
frequency is given to lowest order by
\begin{equation}
  \omega_D = \frac{-i \mu k_0^2 V(k_0)}{\mu k_0^2
  V(k_0) + \tilde \omega_c^2}2|\eta_1|\delta k^2 ,
\end{equation}
where $\delta k = |{\bf k}|-k_0$.  The circulating
current modes to lowest order are
\begin{equation}
  \omega_{\pm} = -i k_0^2 \tilde \eta/2 \pm
  \sqrt{-k_0^4 \tilde \eta^2/4 + \mu _0^2 V(k_0) + \tilde \omega_c^2},
  \label{Bmodes2}
\end{equation}
where $\tilde \eta = \eta_2 + \eta_4 k_0^2$.  We have set
$\omega_0^{-1} = 0$ here since the modes already do not become
critical simultaneously.
In the limit where the $\tilde \omega_c$
term is dominant in Eq.\ (\ref{Bmodes2}), the square root is positive.
We will assume that $\tilde \eta >0$ so that these modes remain damped
when the diffusive mode becomes critical since this appears to be the
experimentally relevant situation.  As the resistance
decreases further, one would expect the
diffusive mode to give rise to a time-independent
state with nonuniform density ordered at wavevector $k_0$.

It follows from the preceding discussion that only the diffusive mode
should be important for describing the transition into a nonuniform
density phase ordered at either zero or finite wavevector.  Since the
circulating current modes have a finite damping rate when the
diffusive mode becomes critical, they can be neglected provided we
focus on frequencies smaller than their decay rate.  This provides a
large simplification in that it allows us to eliminate the currents
altogether and obtain a theory in terms of the density alone.
Physically, this is possible because at long times scales the current
and density fluctuations are dominated by a diffusive mode
characterized by a gradient of the density fluctuations that just
balances the Lorentz force associated with the current fluctuations.
One would thus expect to be able to write the `fast' current in terms
of the `slow' density.  This can be done by dropping the time
derivatives on the left-hand side of Eq.\ (\ref{jeq}) compared to
$\tilde \omega_c$ and then solving order by order for the current as a
function of the density.  Inserting the resulting expression into the
continuity equation yields a decoupled equation of motion for the
density alone.  The transition within this simplified description of
the system will be analyzed in Sections III and IV for the case of
zero wavevector ordering; finite-wavevector ordering is briefly
mentioned in Section V but will not be studied in detail here.

\section{Transition to density-ordered state at zero wavevector with
$\Phi \rightarrow \Phi + const$ symmetry}

When $\eta_1 >0$ so that the resistivity is minimized at zero
wavevector, we saw in the previous section that the zero-current state
of the system becomes unstable when $r < 0$.  Identifying the precise
ordered state that develops in this regime is complicated by the
presence of nonlinear terms in Eq.\ (\ref{jeq}) involving the
magnitude of the potential $\Phi$.  If one ignores such terms by
manually imposing the symmetry $\Phi \rightarrow \Phi + const$, then a
simple ordered state emerges, namely a state with a uniform current
and a transverse electric field that balances the Lorentz force.
We will begin this section by discussing some mean-field properties of this
ordered state and then analyze the transition to this state
using dynamical renormalization group techniques.  Our motivation for
studying this simplified model is as follows.  First, it is the simplest
model that one can construct that captures the instability that
occurs when the resistance becomes negative.  Second, we expect this model
to be appropriate for describing physics on length scales where terms
involving the magnitude of $\Phi$ play a relatively unimportant role.
This will be quantified below.  Third, understanding the properties of
the transition in this minimal model will allow us to better
understand the effects of adding in terms that violate the $\Phi
\rightarrow \Phi + const$ symmetry, which will be done in Section IV.

\subsection{Ordered state and linearized theory of fluctuations}

When $r<0$, the ordered state within a model with $\Phi \rightarrow
\Phi + const$ symmetry consists of a uniform current
\begin{equation}
  {\bf j}_0 =
  \sqrt{|r|/u} {\bf \hat x},
  \label{constj}
\end{equation}
where the direction ${\bf \hat x}$ is spontaneously picked out.  To
balance the associated Lorentz force requires an electric field given
by
\begin{equation}
  {\bf E}_0 = -\nabla \Phi_0 = (\tilde
  \omega_c/\mu){\bf j}_0\times {\bf \hat z}.
  \label{E2}
\end{equation}
We have assumed here that $j_0$ is small in some sense so that,
for instance, terms
in the equation of motion proportional to $|{\bf j}|^4 {\bf j}$ can be
neglected compared to the $u|{\bf j}|^2{\bf j}$ term.  To further
simplify things, terms such as $(\nabla \Phi)^2 {\bf j}$ that would
arise from expanding the longitudinal resistivity to higher order in
the potential have also been neglected.  Their presence only alters
quantitative
properties of the ordered state.  For instance, a uniform current
still develops, but with a modified magnitude.

To establish a connection with the experiments, note that the
longitudinal resistance at zero wavevector and frequency is
proportional to $-|r|+u|{\bf j}|^2$ (neglecting higher-order terms in
${\bf j}$ and $\nabla \Phi$).  The spontaneous current $j_0$ therefore
corresponds to a vanishing longitudinal resistance as seen
experimentally.  This is also consistent with the results of Andreev
\emph{et al}.\cite{Andreev} that show that a stable state must have
spontaneous currents corresponding to a vanishing longitudinal
resistivity.

To analyze the stability of the ordered state,
we consider fluctuations about the uniform current state by writing ${\bf j}
= {\bf j}_0 + {\bf \delta j}$ and $n = n_0 + \delta n$, where
${\bf j}_0$ is given in Eq.\ (\ref{constj}) and $n_0$ corresponds to
the potential $\Phi_0$ given in Eq.\ (\ref{E2}).  In the
linearized equations of motion for ${\bf \delta j}$ and $\delta n$,
there are two damped modes in the $k\rightarrow 0$ limit with frequencies
\begin{equation}
  \omega_{\pm}' = -i|r|\pm \sqrt{\tilde \omega_c^2-|r|^2}.
\end{equation}
Since the ordered state breaks rotational symmetry, there is also
a Goldstone mode with frequency $\omega_G$ whose real and imaginary
parts are given by
\begin{eqnarray}
  \text{Re} \omega_G &=& \frac{\mu j_0 \nu_1}{\tilde\omega_c^2}
  k_{\parallel} k^2 V(k)
  \\
  \text{Im} \omega_G &=& -\frac{\mu}{\tilde\omega_c^2}(2 u j_0^2
  k_\perp^2 + \eta_1 k_\parallel^4)V(k),
\end{eqnarray}
where $k_\perp$ and $k_\parallel$ are the components of ${\bf k}$
perpendicular and parallel to ${\bf j}_0$, respectively.
Note that the damping within this mode is anisotropic.  In particular,
fluctuations with wavevector parallel
to ${\bf j}_0$, which produce long-wavelength variations in the
direction of ${\bf j}_0$, relax much more slowly than fluctuations
with wavevector perpendicular to ${\bf j}_0$.

To see if the ordered state is stable to fluctuations, one needs to
calculate the mean-squared fluctuations of $\delta {\bf j}({\bf r},t)$
and $\delta n({\bf r},t)$, averaged over the noise.  A divergence of
either of these quantities would signal the destruction of the ordered
state.  We compute these quantities within the linearized theory,
focusing only on
fluctuations arising from the Goldstone mode for simplicity.
(Long-wavelength fluctuations arising from the $\omega'_{\pm}$
modes will be finite since they have a nonzero damping rate
as $k \rightarrow 0$; consequently, these modes can be neglected.)
Denoting the current fluctuations parallel and perpendicular
to ${\bf j_0}$ by  ${\bf \delta j}_{\parallel}$ and ${\bf \delta j}_{\perp}$,
respectively, we find
\begin{eqnarray}
\!\!\!\!\!\!  \langle {\bf \delta j}^2_\parallel({\bf r},t) \rangle \!\!\! &\approx&\!\!\!
  \frac{g}{\tilde \omega_c^4} \int_{\bf k} \frac{\mu V(k) k_\perp^2
  (\tilde \omega_c^2 k^2 - \alpha^2 k_\perp^2)}{\alpha k_\perp^2 +
  \eta_1 k^4}
  \label{jpara} \\
\!\!\!\!\!\!  \langle {\bf \delta j}^2_\perp({\bf r},t) \rangle \!\!\! &\approx& \!\!\!
  \frac{g}{ \tilde \omega^6} \int_{\bf k} \frac{\mu V(k)[\tilde
  \omega_c^4 k^2 k_\parallel^2 - \alpha^2 (\tilde \omega_c^2 + \alpha^2)
  k_\perp^4]}{\alpha k_\perp^2 + \eta_1 k^4},
  \label{jperp}
\end{eqnarray}
where $\alpha = 2 u j_0^2$ and $g$ is the noise strength.  Equation
(\ref{jpara}) is obviously finite with either short-range interactions
($V(k) \sim const$) or long-range interactions ($V(k) \sim 1/k$) since
the integrand itself is not infrared divergent.  With long-range
interactions, the integrand in Eq.\ (\ref{jperp}) is infrared
divergent. However, this divergence is integrable in 2D, leading to
finite transverse current fluctuations.  The mean-squared density
fluctuations are given by
\begin{equation}
  \langle \delta n^2({\bf r},t)\rangle \approx \frac{g}{\tilde
  \omega_c^2} \int_{\bf k}\frac{\tilde \omega_c^2 k^2 + \alpha^2 k_\perp^2
  - 2 \alpha \tilde \omega_c
  k_\parallel k_\perp}{\mu V(k)(\alpha k_\perp^2 + \eta_1 k^4)}.
  \label{nfluct}
\end{equation}
Again, since the infrared divergence in Eq.\ (\ref{nfluct}) is
integrable with either short- or long-range interactions, the density
fluctuations are also finite.  Hence we conclude that, for
sufficiently low noise, $g$, the spontaneous current-carrying state is
stable to current and density fluctuations with either short- or
long-range interactions.

The state characterized by Eqs.\ (\ref{constj}) and (\ref{E2}) can
clearly not exist in arbitrarily large samples since the density would
eventually become negative on one side of the sample. For a given
spontaneous current ${\bf j}_0$, one can estimate the maximum sample
length $L_{\rm c1}$ below which this is a sensible ordered state by
finding how large the sample can be before the density change becomes
comparable to the mean density.  We do this by assuming that the
electron-electron interactions are screened so that the electric field
is given by the gradient in the electrochemical potential $\mu$.  If
$\Delta \mu$ is the change in $\mu$ between the edges of a sample of
length $L$, then the magnitude of the electric field is $E = \Delta
\mu/eL$, where $e$ is the electron charge.  Regions of density
variation comparable to the mean density will appear if $\Delta \mu
\sim E_F$, where $E_F$ is the Fermi energy.  Setting $\Delta \mu =
E_F$ and using Eq.\ (\ref{E2}), we get
\begin{equation}
  L_{\rm c1} \sim \frac{E_F}{e\rho_H j_0},
  \label{Lc}
\end{equation}
where we have identified $\tilde \omega_c /\mu = \rho_H$.

Note that as the mean-field critical point is approached, $j_0
\rightarrow 0$ and so $L_{\rm c1}$ diverges.
One might therefore be tempted to conclude that the model with $\Phi
\rightarrow \Phi + const$ symmetry correctly describes the physics at
the transition at all length scales.  We stress that this is not
necessarily the case.  In computing $L_{\rm c1}$, we have only
demanded that no unphysical features such as negative density arise in
this minimal model.  What we have \emph{not} done is compute the
characteristic length (which can be smaller than $L_{\rm c1}$ above)
below which terms that depend on the magnitude of $\Phi$ play a
negligible role.  We will elaborate further on this in the following
subsection.

We now estimate $L_{\rm c1}$ using parameters measured by Willett
\emph{et al}.\cite{Willett} in order to get a feel for this length
scale. In their experiments, carried out on GaAs/AlGaAs samples, the
density is $n \approx 2\times 10^{11} {\rm cm}^{-2}$, from which we
estimate $E_F \sim 5$meV. In a 20GHz microwave field the primary zero
resistance state occurs at $B \approx 0.4$kG, where $\rho_H=B/ne
\approx 125 \Omega$.  From spontaneous voltages that develop in this
zero resistance region, they estimate a spontaneous current of roughly
$5 \mu$A flowing between the center and edge in square samples of
length $0.4$mm. Assuming a single domain between these contacts,
we find $j_0 \approx 25 \mu
\text{Amm}^{-1}$.   Putting these parameters together, we estimate the critical length for
to be $L_{\rm c1} \sim 1$mm.  In samples with dimension larger than
$L_{\rm c1}$, terms in the equation of motion involving the magnitude
of $\Phi$ must be taken into account to produce a sensible ordered
state. Such terms would prevent the density from becoming arbitrarily
large and negative, and would lead to inhomogeneous currents and
densities. Determining the corresponding current-carrying ordered
state on these longer length scales is an interesting problem that we
do not address here.

\subsection{Transition with short-range interactions}

Having discussed an example of a stable ordered state that arises from
a model with $\Phi \rightarrow \Phi + const$ symmetry when $r<0$, we
now turn to the critical properties of system at the phase transition.
We begin with the simplest case of short-range interactions.

As discussed in Section II, our analysis is greatly simplified by
assuming that near the critical point circulating current modes remain
damped while the diffusive mode become critical.  Focusing only on the
diffusive mode enables us to write the current in terms of the density
and use the continuity equation to write an effective theory involving
only the density.  Imposing the symmetry $\Phi\rightarrow \Phi +
const$, the resulting equation of motion takes the form
\begin{eqnarray}
  0 &=& \partial_t n -\tilde r \nabla^2 \Phi+ D \nabla^4 \Phi -\nabla \cdot {\bf \zeta'}
  + \lambda_1 \nabla \cdot (\nabla^2\Phi\nabla\Phi)
  \nonumber \\
  &+& \lambda_2{\bf \hat z} \cdot
  (\nabla\Phi\times\nabla^3\Phi)
  +\lambda_3\nabla\cdot(({\bf \hat z}\times\nabla\Phi)\cdot\nabla({\bf
  \hat z}\times \nabla\Phi))
  \nonumber \\
  &+& \mu_1\nabla\cdot(\nabla\Phi(\nabla\Phi)^2)
  + \mu_2{\bf \hat z}
  \cdot(\nabla\Phi\times\nabla(\nabla\Phi)^2),
  \label{Phi1}
\end{eqnarray}
where $D>0$ and ${\bf \zeta'}$ is a Gaussian noise source with
variance $2g'$.  The transition occurs at $\tilde r = 0$ (within
mean-field theory), so we will take $\tilde r = 0$ from now on.  Since
we are considering short-range interactions here, $\Phi[n] = n$.
Quartic and higher order terms in $\nabla\Phi$ have been neglected
since they contain at least five gradient operators and are therefore
irrelevant in two dimensions.

The upper critical dimension for this model is $d_{uc} = 2$ for the
case of short-range interactions.  We use dynamical renormalization
group techniques\cite{RGref1,RGref2} to deduce whether the
nonlinearities appearing in Eq.\ (\ref{Phi1}) are marginally relevant
or irrelevant in two dimensions.  This procedure is facilitated by the
use of the Martin-Siggia-Rose (MSR) formalism\cite{MSR}.
The essence of this formalism is that one introduces a `partition
function' $Z$ that is useful for obtaining various correlation
functions, namely
\begin{equation}
  Z = \int\mathcal{D}n\delta(\partial_t n - \tilde r\nabla^2 \Phi + D
  \nabla^4 \Phi - \nabla \cdot {\bf \zeta} +
  \cdots). \label{partition1}
\end{equation}
This imposes the equation of motion as a constraint on all possible
spacetime `trajectories' of $n(\br,t)$. The ellipsis indicates all
nonlinearities appearing in Eq.~(\ref{Phi1}). This functional
delta-function constraint is implemented through an auxiliary field
$\tilde{n}(\br,t)$ so that $Z$ can be written as
\bea
  Z &=& \int \mathcal{D}n \mathcal{D}\tilde{n} e^{i S[n,\tilde n]} \\
  S &=& \int_{{\bf r},t}\tilde n[\partial_t n -\tilde r \nabla^2 \Phi
  + D \nabla^4 \Phi - \nabla \cdot \vec \zeta' + \cdots],
\eea
where constants have been absorbed into the integration measure for
$\tilde n$. A useful feature of this method is that the noise
averaging can now be easily performed, with the result that
\begin{equation}
  \int_{{\bf r},t}\tilde n[-\nabla \cdot \zeta']
  \rightarrow ig'\int_{{\bf r},t}(\nabla\tilde n)^2
  \label{noiseave}
\end{equation}
in the `action' $S$.
This leads to an MSR action expressed
in terms of the fields $n,\tilde{n}$ and no noise terms.
One can then implement
the renormalization group transformation using standard field
theory techniques as follows.
First, the
action $S$ is written in Fourier space with an ultraviolet cutoff
$\Lambda$ reflecting the coarse-graining of the fields.
One then integrates out fields with wavevectors $q$ such that
$\Lambda/s < q <\Lambda$, where $s > 1$.  This results in an effective
action with a reduced cutoff $\Lambda/s$.  To restore the initial
cutoff, one then rescales wavevectors, frequencies, and the fields
according to
\begin{eqnarray}
  k' &=& sk
  \\
  \omega' &=& s^z \omega
  \\
  n'({\bf k}',\omega') &=& s^{-\chi}n({\bf k},\omega)
  \\
  \tilde n'({\bf k}',\omega') &=& s^{-\tilde \chi}\tilde n({\bf k},\omega).
\end{eqnarray}
By setting $s = 1 + d\ell$, one then obtains differential recursion
relations that specify how the effective coupling constants for the
long-scale degrees of freedom ``flow'' as short-scale degrees of
freedom are integrated out. In the present paper these recursion
relations will be calculated to one-loop order.

In anticipation of finding a stable Gaussian fixed point, we choose
the rescaling exponents to take on their mean-field values: $z = 4$,
$\chi = 6$, and $\tilde \chi = 4$.  (Since there is no small parameter
at our disposal, the only possible controlled fixed point
\emph{must} be Gaussian.)  These exponents keep the noise strength
$g'$ fixed under renormalization since diagrammatic corrections to
$g'$ vanish at one-loop order.  To simplify the flow equations for
the remaining coupling constants, we define the following
dimensionless parameters:
\begin{eqnarray}
  \Lambda_1 &=& (\alpha/D) \lambda_3 (4\lambda_1 - 3\lambda_3)
  \nonumber \\
  \Lambda_2 &=& -(\alpha/D)\lambda_3\lambda_2
  \nonumber \\
  \Lambda_3 &=& (\alpha/D) \lambda_3^2
  \label{Lambdas} \\
  \Lambda_4 &=& -\alpha \mu_1
  \nonumber \\
  \Lambda_5 &=& \alpha \mu_2,
  \nonumber
\end{eqnarray}
where $\alpha = g/4\pi D^2$.
The flow equations in terms of these parameters are
\begin{eqnarray}
  \partial_\ell D &=& \frac{1}{2}\Lambda_1 D
  \label{D} \\
  \partial_\ell \Lambda_1 &=& -(\frac{3}{2} \Lambda_1+13\Lambda_4)\Lambda_1
  - 3\Lambda_3 \Lambda_4 + 4 \Lambda_2 \Lambda_5
  \label{L1} \\
  \partial_\ell \Lambda_2 &=& -(\frac{3}{2} \Lambda_1 + 7 \Lambda_4)\Lambda_2
  + \frac{1}{4}(3 \Lambda_3 - 7\Lambda_1) \Lambda_5
  \label{L2} \\
  \partial_\ell \Lambda_3 &=& -(\frac{3}{2} \Lambda_1 + 12\Lambda_4) \Lambda_3
  \label{L3} \\
  \partial_\ell \Lambda_4 &=& -(\Lambda_1 + 9 \Lambda_4)\Lambda_4 +
  \Lambda_5^2
  \label{L4} \\
  \partial_\ell \Lambda_5 &=& -(\Lambda_1 + 10 \Lambda_4)\Lambda_5.
  \label{L5}
\end{eqnarray}

At this point we would like to identify the basin of attraction for
the Gaussian fixed point under consideration.  That is, for a given
set of initial conditions for $\Lambda_i$, we would like to know
whether these parameters all flow to zero as $\ell \rightarrow
\infty$.  While it is straightforward to check this numerically,
it is difficult to draw general conclusions either analytically or 
from the numerics due to the five-dimensional parameter space 
and the fact that Eqs.\ (\ref{L1})
through (\ref{L5}) are all coupled.  In the subspace with 
$\lambda_3 = 0$, one can show analytically that the Gaussian fixed
point is stable to all perturbations within that subspace.  
In the full parameter space with $\lambda_3 \neq 0$, we have shown that a
finite-volume region of initial conditions corresponds to stable
trajectories where each $\Lambda_i$ flows to zero.  The asymptotic
solution for such trajectories is given by $\Lambda_1 \sim (2/11)
\ell^{-1}$, $\Lambda_2 \sim (5 c_2 /242) \ell^{-10/11}$, $\Lambda_3
\sim c_1 \ell^{-15/11}$, $\Lambda_4 \sim (1/11)\ell^{-1}$, and
$\Lambda_5 \sim (1/c_2) \ell^{-12/11}$, where $c_{1,2}$ are arbitrary
constants.  One can verify the stability of these flows by perturbing
around this solution.  According to Eq.\ (\ref{D}), the subdiffusion
constant grows asymptotically as $D \sim D_0\ell ^{1/11}$ along these
trajectories, where $D_0$ is a constant.  The asymptotic behavior of
the original coupling constants is given by $\lambda_1 \sim
\ell^{-2/11}$, $\lambda_2 \sim \ell^{-1/11}$, $\lambda_3 \sim
\ell^{-6/11}$, $\mu_1 \sim \ell^{-9/11}$, and $\mu_2 \sim
\ell^{-10/11}$, demonstrating marginal irrelevance of all nonlinearities,
and therefore the stability of the Gaussian fixed point.

Flows that terminate along the above asymptotic trajectories
correspond to marginally irrelevant couplings that reside in the basin
of attraction for the Gaussian fixed point.  In such cases mean-field
theory should be a reasonable starting point for analyzing the
transition to the density-ordered phase within this model.  In
particular, as we saw in the previous subsection mean-field theory
predicts a \emph{continuous} transition since the spontaneous current
and the associated density gradient develop smoothly from zero as $r$
becomes negative (see Eqs.\ (\ref{constj}) and (\ref{E2})).  Another
important mean-field prediction we can make is that at the transition
there is a single subdiffusive mode for density fluctuations with
frequency
\begin{equation}
  \omega = -i D k^4.
  \label{slowmode}
\end{equation}
This slow relaxation of long-wavelength fluctuations should be
accompanied by large voltage fluctuations near the transition.
Equal-time density-density correlations, which should mimic voltage
correlations, are given within mean-field theory by
\begin{equation}
  \langle n({\bf r},t) n({\bf 0},t) \rangle = \frac{g'}{D}
  \int_{\bf k} \frac{e^{i {\bf k}\cdot{\bf r}}}{k^2}.
  \label{ncorr}
\end{equation}
The integral diverges logarithmically at small $k$.  To regulate the
integral, we restrict the range of integration to $2\pi/L < k
<\Lambda$, where $L$ is the system size.  In the limit $r\Lambda \gg
1$ and $r/L \ll 1$, we obtain
\begin{equation}
  \langle n({\bf r},t) n({\bf 0},t) \rangle \approx \frac{g'}{2\pi D}
  \ln(L/r).
\end{equation}
Equal-time current-current correlations in mean-field theory are given by
\begin{equation}
  \langle {\bf j}({\bf r},t){\bf j}({\bf 0},t) \rangle =
  \frac{g'\mu\Lambda^2}{2 \pi D \tilde
  \omega_c^2}\frac{J_1(r\Lambda)}{r\Lambda},
  \label{jcorr}
\end{equation}
where $J_1(x)$ is a Bessel function of the first kind.  Note that the
current becomes $\delta$-function correlated in the limit that
$\Lambda
\rightarrow \infty$.  Since the interactions are only marginally
irrelevant, they will give rise to logarithmic corrections to these
correlation functions, which will not be computed here.

So far we have focused only on the case where the coupling constants
flow to zero upon renormalization.  Even outside of this marginally
stable region, the couplings are still only marginally relevant, and
therefore grow only logarithmically with length scale.  In fact, we
have seen numerically that many trajectories that initially flow
toward the Gaussian fixed point eventually diverge from it, but
do so only after many renormalization group iterations.  In these
instances, it may be very difficult to resolve deviations from
mean-field theory either numerically or experimentally and the
transition may appear continuous even in the presence of the
marginally relevant couplings.

\subsection{Is this model valid near the transition?}

We will now discuss when the model with $\Phi \rightarrow \Phi +
const$ symmetry and short-range interactions is expected to be
appropriate for describing the physics at the transition.  Consider
adding the term $\lambda \nabla\cdot(\Phi\nabla \Phi)$ to Eq.\
(\ref{Phi1}), which is the most relevant nonlinearity that violates
this symmetry. This term can be traced back to the $\gamma_2 \Phi {\bf
j}$ term in Eq.\ (\ref{jeq}).  We define a dimensionless coupling
constant $\tilde \lambda \equiv \lambda/D$, where $D$ is the
subdiffusion constant. If we interpret the equation of motion as
arising from a Taylor expansion of the resistivity, then we can write
\begin{equation}
  \tilde \lambda = \frac{\partial \rho_D/\partial
  n}{\partial\rho_D/\partial k^2}.
\label{micro}
\end{equation}
One expects $\tilde \lambda$ to be small in a not-too-dirty electron
gas, since in a pure system $\rho_D$ is already non-vanishing at any
nonzero wavevector (contributing to the denominator), while the
numerator vanishes by Galilean invariance in this limit. However, this
term is strongly relevant in two dimensions. Hence, even if one starts
with $\tilde \lambda \ll 1$, in an infinite system this coupling
constant will eventually become much greater than unity under
renormalization. Ignoring this term will certainly not be valid in
this case, so one would need to appeal to the full equation of motion
to describe the transition. In a finite system, however, one is
interested in reducing the cutoff to roughly $1/L$, where $L$ is the
system size, so the growth of the coupling constant will be
bounded. The model with $\Phi \rightarrow \Phi + const$ symmetry will
provide a reasonable description of the transition as long as $L$ is
sufficiently small that $\tilde \lambda$ does not become of order
$1$. Under a tree-level renormalization group iteration, the
renormalized coupling constant $\tilde \lambda'$ grows according to
$\tilde \lambda' = \tilde \lambda s^2$, with $s>1$.  In terms of the
reduced cutoff $\Lambda' = \Lambda/s$, where $\Lambda$ is the initial
cutoff, this can be expressed as
\begin{equation}
  \frac{\tilde \lambda'}{\tilde \lambda} =
  \bigg{(}\frac{\Lambda}{\Lambda'}\bigg{)}^2.
\end{equation}
We take $\Lambda' = 1/L_{\rm c2}$ and $\Lambda = 1/l_{\rm in}$, where
$l_{\rm in}$ is the inelastic mean free path. For the samples used in
the experiments, $l_{\rm in}\sim \hbar v_F E_F / (k_B T)^2
\sim 100 \mu$m\cite{Dmitriev1}, and is comparable to the transport mean
free path estimated from the mobility at a temperature of
1K.  This
is about an order of magnitude smaller than the sample lengths.

To estimate the critical length scale $L_{\rm c2}$ below
which the model is valid, we set $\tilde \lambda' = 1$, leading to
\begin{equation}
L_{\rm c2} = \frac{l_{\rm in}}{\sqrt{|\tilde \lambda}|}.
\end{equation}
We note that if $|\tilde{\lambda}| \ll 1$,
say around $0.01$, then the
critical length $L_{\rm c2}$ would already be comparable to the
sample sizes studied in the experiments.
A serious estimate of this length would require a microscopic
calculation of the resistivity $\rho_D(\bk,\omega)$
in the presence of microwaves to
compute the bare value of $\tilde \lambda$ via Eq.~(\ref{micro})
and would be valuable.

\subsection{Transition with long-range interactions}

We have seen in the case of short-range interactions above that a
finite-volume region of initial couplings are marginally irrelevant
and flow to zero upon coarse-graining.  Next, we discuss the fate of
these flows when long-range interactions are turned on.  This case is
relevant experimentally due to the absence of metallic gates in the
experiments conducted so far, leading to unscreened Coulomb
interactions.

Consider again the equation of motion given in Eq.\ (\ref{Phi1}), with
$\Phi[n] = \int_{{\bf r}'} V({\bf r-r}')n({\bf r}')$.  With long-range
Coulomb interactions, the Fourier-transformed interaction potential
(in two dimensions) is
$V(k) \sim 1/k$.  The upper critical dimension in this case is $d_{uc}
= 3$.  Since we are interested in the transition in $d = 2$
dimensions, one option is to carry out an $\epsilon$-expansion in
$d = 3-\epsilon$ dimensions.  This approach is complicated by the need
to generalize the interactions in Eq.\ (\ref{Phi1}) to higher dimensions.
Alternatively, one can perform an $\epsilon$-expansion by writing
$V(k) = 1/k^{\epsilon}$, with $\epsilon \ll 1$.
The upper critical dimension is then $d_{uc} = 2 + \epsilon$.  We will
adopt the latter approach since we can then work directly in $d = 2$
dimensions and thereby avoid generalizing the equation of motion.

We use the dynamical renormalization group as outlined above to
calculate the flow equations at one-loop order and to lowest order in
$\epsilon$.  As in the short-range case, there are no diagrammatic
corrections to the noise strength $g'$ at one-loop.  To keep $g'$
fixed under renormalization, we take the rescaling exponent $\tilde
\chi = (4 + z)/2$.  Similarly, we choose the exponent $\chi = 3z/2$ to
fix the coefficient of $\partial_t n$ to be unity in Eq.\
(\ref{Phi1}).  To simplify the flow equations for the remaining
parameters, we again use the dimensionless coupling constants defined
in Eq.\ (\ref{Lambdas}) (with $\alpha = g/4\pi D^2
\Lambda^{\epsilon}$).  The subdiffusion constant then flows
according to
\begin{equation}
  \partial_l D = (z-4 + \epsilon + \Lambda_1/2) D.
\end{equation}
For convenience we choose $z = 4 - \epsilon - 1/2 \Lambda_1$ to keep
$D$ fixed.

With this choice of rescaling exponents, the flow equations for the
parameters $\Lambda_i$ are
\begin{eqnarray}
\!\!\!\!\!\!\!\!  \partial_\ell \Lambda_1 &=& \epsilon \Lambda_1 -(\frac{3}{2} \Lambda_1+13\Lambda_4)\Lambda_1
  - 3\Lambda_3 \Lambda_4 + 4 \Lambda_2 \Lambda_5
  \label{L1b} \\
\!\!\!\!\!\!\!\!  \partial_\ell \Lambda_2 &=& \epsilon \Lambda_2 -(\frac{3}{2} \Lambda_1 + 7 \Lambda_4)\Lambda_2
  + \frac{1}{4}(3 \Lambda_3 - 7\Lambda_1) \Lambda_5
  \label{L2b} \\
\!\!\!\!\!\!\!\!  \partial_\ell \Lambda_3 &=& \epsilon \Lambda_3 -(\frac{3}{2} \Lambda_1 + 12\Lambda_4) \Lambda_3
  \label{L3b} \\
\!\!\!\!\!\!\!\!  \partial_\ell \Lambda_4 &=& \epsilon \Lambda_4 -(\Lambda_1 + 9 \Lambda_4)\Lambda_4 +
  \Lambda_5^2
  \label{L4b} \\
\!\!\!\!\!\!\!\!  \partial_\ell \Lambda_5 &=& \epsilon \Lambda_5 -(\Lambda_1 + 10 \Lambda_4)\Lambda_5.
  \label{L5b}
\end{eqnarray}
In the case of short-range interactions we found that there are stable
trajectories where all the coupling constants go asymptotically to
zero.  This clearly cannot happen in the case of finite-range
interactions due to the $\epsilon \Lambda_i$ terms above.  Instead, we
search for fixed points of the form $\Lambda_i = a_i \epsilon$, where
$a_i$ are constants.  One can easily show that all such fixed points
are unstable.  We interpret this lack of a stable fixed point as
signaling a first-order transition.  Thus, we conclude that \emph{the
continuous transition that can occur with short-range interactions is
driven first-order by the presence of long-range interactions of the
form $V(k) = 1/k^{\epsilon}$, with $\epsilon \ll 1$}.

This result may seem surprising initially since one might expect
long-range interactions to suppress density fluctuations and thereby
further stabilize the Gaussian fixed point.  For instance, in the
linearized equation of motion the density diffuses faster with
long-range interactions.  A competing effect, however, is that density
fluctuations can interact nonlocally through the nonlinear terms.
Thus, density fluctuations in one region of the sample can further
induce fluctuations over long distances.  This can lead to positive
feedback of these density fluctuations via the nonlinearities, which
evidently drives the transition first-order.

\section{Transition to density-ordered state at zero wavevector in the
full rotationally invariant model}

The model considered above with $\Phi \rightarrow \Phi + const$
symmetry is only appropriate for describing physics up to a certain
length scale.  For instance, in the ordered state with a given uniform
current, regions of negative density appear if the sample is too
large.  We estimate this length scale to be roughly $\sim 1$mm using
parameters from Willett's experiments.\cite{Willett} On larger scales,
terms in the equation of motion depending on the magnitude of $\Phi$,
which prevent the density from becoming negative, must be taken into
account.  As discussed above, at the transition, the leading term
involving the magnitude of $\Phi$ (i.e., the $\gamma_2 \Phi {\bf j}$
term in Eq.\ (\ref{jeq})) is strongly relevant in two dimensions.  The
dimensionless coupling constant for this term therefore grows under
renormalization.  In a finite system, the growth of this coupling is
limited by the system size $L$ since one only reduces the wavevector
cutoff to of order $1/L$.  Neglecting terms depending on the magnitude
of $\Phi$ becomes an invalid approximation when the system size is
sufficiently large that this renormalized dimensionless coupling
becomes of order unity.

To describe physics in samples with linear dimensions larger than
these length scales, one must therefore relax the $\Phi \rightarrow
\Phi + const$ symmetry and appeal to the full equation of motion in
Eq.\ (\ref{jeq}) with no additional symmetries.  This is the subject
of the present section.  Identifying the ordered state that develops
in this case is nontrivial, so we will focus only on the transition to
the ordered state, considering both short- and long-range interactions.

\subsection{Transition with short-range interactions}

When we relax the $\Phi \rightarrow \Phi + const$ symmetry, Eq.\
(\ref{Phi1}) generalizes to
\begin{eqnarray}
  0 = \partial_t n - \tilde r\nabla^2\Phi + D \nabla^4 \Phi - \nabla
  \cdot {\bf\zeta'}- \lambda \nabla \cdot(\Phi\nabla\Phi),
  \label{Phi2}
\end{eqnarray}
where $D>0$ and ${\bf \zeta'}$ is a Gaussian noise source with
variance $2g'$.  In this subsection we consider short-range
interactions, so $\Phi = n$.  The transition in the linearized theory
occurs at $\tilde r = 0$ since the diffusive mode becomes unstable
when $\tilde r <0$.
Other nonlinearities are in principle present in Eq.\
(\ref{Phi2}), but they are less
relevant than the $\lambda$ term and can be neglected provided we work
near the upper critical dimension.

To derive Eq.\ (\ref{Phi2}), we solved for the current in terms of the
density assuming two spatial dimensions.  However, the $\lambda$
interaction is strongly relevant in two dimensions, so to study its
effects we need to to continue this model to higher dimensions.  We
initially adopt the most naive way of doing this continuation; namely,
we simply assert that Eq.\ (\ref{Phi2}) is valid in $d$ dimensions.
In the case of short-range interactions the upper-critical dimension
for the $\lambda$ nonlinearity is then $d_{uc} = 6$.

\begin{figure}
  \begin{center}
    {\resizebox{8cm}{!}{\includegraphics{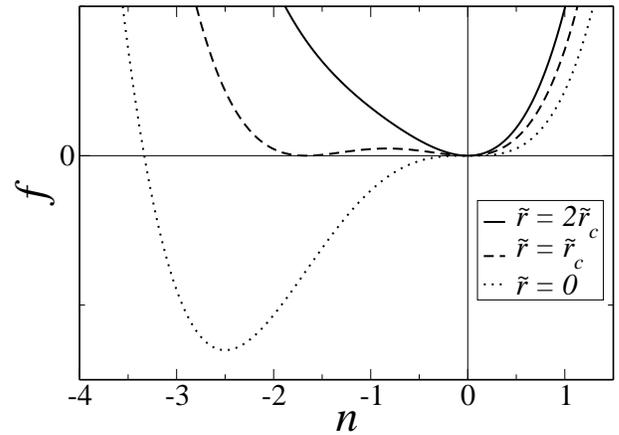}}}
  \end{center}
  \caption{``Free energy'' density $f$ as a function of constant
    density $n$ for
    $\lambda = 1$, $\tilde u = 1/5$, with $\tilde r = 2 \tilde r_c$,
    $\tilde r_c$, $0$ (see text for details). }
  \label{fig2}
\end{figure}

We have carried out an $\epsilon$-expansion in $d = 6-\epsilon$
dimensions to obtain the renormalization group flow equations to
one-loop order.  Rather than go through the details of the
calculation, we will merely state that these equations lack a stable
fixed point, which we interpret as signaling a first-order transition.
A more direct route to this conclusion can be obtained by observing
that Eq.\ (\ref{Phi2}) is identical to the equation of motion for an
\emph{equilibrium} model with a conserved order parameter.  That is, 
it (Eq.\ (\ref{Phi2})) can be rewritten as
\begin{equation}
  \partial_t n= \nabla^2\frac{\delta F}{\delta n} + \nabla\cdot {\bf
  \zeta'}, \label{eqn}
\end{equation}
with the ``free energy'' given by
\begin{equation}
  F = \frac{1}{2}\int_r\bigg{[}\tilde r n^2+D(\nabla n)^2 +
  \frac{\lambda}{3} n^3 + \frac{\tilde u}{2}n^4\bigg{]},
\end{equation}
The term proportional to $\tilde u$ that results from Eq.\ (\ref{eqn})
is irrelevant at the upper critical dimension and has therefore been
excluded from Eq.\ (\ref{Phi2}).  We will assume $\tilde u >0$ for
simplicity, although this is not essential.  Figure \ref{fig2} depicts
the free energy density $f$ as a function of uniform density $n$ for
three different values of $\tilde r$.  When $\tilde
r>\lambda^2/18\tilde u\equiv \tilde r_c$, the free energy is minimized
when $n = 0$ as illustrated by the solid curve.  At $\tilde r = \tilde
r_c$, the free energy has two degenerate minima as shown in the dashed
curve.  Below $\tilde r_c$, the free energy is minimized by a nonzero
value of $n$.  This situation is represented by the dotted line for
$\tilde r = 0$.  When $\tilde r$ decreases below $\tilde r_c$, the
density will therefore jump discontinuously from zero to minimize the
``free energy''.  This signals the onset of a first-order transition,
consistent with our renormalization group results.  Note that the
transition occurs at a finite value of $\tilde r$, preempting the
apparent transition (a `spinodal') at $\tilde r = 0$ expected from the
linear theory.  As Fig.\ \ref{fig2} demonstrates, the point $\tilde r
= 0$ actually corresponds to a spinodal decomposition where the system
goes from being metastable to globally unstable at $n = 0$.

These results hold only near $d = 6$.  We can reduce the upper
critical dimension of the model by considering a spatially anisotropic
continuation of Eq.\ (\ref{Phi2}) to higher dimensions.  To do this,
we can start by continuing Eq.\ (\ref{jeq}) to $d$ dimensions and
taking the resistance at zero wavevector and frequency to be
anisotropic.  That is, write
\begin{equation}
  r{\bf j} \rightarrow  r_\perp {\bf j}_\perp + r_\parallel {\bf j}_\parallel
\end{equation}
in Eq.\ (\ref{jeq}), where ${\bf j}_\perp$ is the current in the
$x$-$y$ plane and ${\bf j}_\parallel$ represents the current in the
additional $d-2$ dimensions.  We will be interested in tuning the
resistance $r_\perp$ in the $x$-$y$ plane to zero while leaving the
resistance $r_\parallel$ for the remaining directions positive.  We
can then eliminate the current in favor of the density as before to
obtain
\begin{eqnarray}
  0 &=& \partial_t n - \tilde r_\parallel \nabla_\parallel^2\Phi
  + D_\perp \nabla_\perp^4 \Phi
  \nonumber \\
  &-& \nabla_\perp \cdot {\bf\zeta_\perp'}-
  \lambda_\perp \nabla_\perp \cdot(\Phi\nabla_\perp\Phi),
  \label{Phi3}
\end{eqnarray}
where $\tilde r_\parallel, D_\perp >0$ and we have set $r_\perp = 0$.
We have also only retained the $x$-$y$ (in-plane) components
of the noise ${\bf \zeta_\perp'}$, as noise components in the
additional $d-2$ $\parallel$ dimensions are irrelevant.  Similarly, we
have omitted nonlinear terms involving $\nabla_\parallel$ since,
due to the high anisotropy of the harmonic terms, these are clearly less
relevant than the corresponding terms involving only $\nabla_\perp$
derivatives.

The upper critical dimension for this model is $d_{uc} = 4$.  We have
performed an $\epsilon$-expansion in $d = 4 - \epsilon$ dimensions to
one-loop order.  Once again, we find that the model lacks a stable
fixed point.  Thus, even near $d = 4$ dimensions, the transition still
appears to be first-order.

It seems quite likely that the transition is first-order in $d = 2$
dimensions as well.  In the model with $\Phi \rightarrow \Phi + const$
symmetry and short-range interactions, we showed in the previous
section that one \emph{could} have a continuous transition in $d = 2$
dimensions.  Terms that violate the $\Phi \rightarrow \Phi + const$
symmetry appear then to always drive the transition first-order.  

We propose the following simple physical interpretation for this.  We
have been analyzing the transition at zero wavevector and zero
frequency, where one expects long-wavelength fluctuations that become
critical at the transition to give rise to an ordered state with
uniform, static current.  Such an ordered state must be accompanied by
a density gradient transverse to the current to balance the Lorentz
force.  We have already argued that such a state cannot exist in the
thermodynamic limit because the density would become arbitrarily large
and negative at the edges of the sample.  The only terms in the
equation of motion that sense these unphysical features are precisely
those terms that depend on the magnitude of the density.  In the
thermodynamic limit, these terms must therefore induce a first-order
transition into some other state, such as a state ordered at finite
wavevector or a phase-separated state.  A direct transition from a
uniform isotropic liquid to a modulated (finite wavevector) smectic
state can also be argued to be first-order on quite general
grounds.\cite{note, Alexander, Brazovskii}

\subsection{Transition with long-range interactions}

Finally, let us consider the effect of long-range interactions.  We
saw in the model with $\Phi \rightarrow \Phi + const$ symmetry that
turning on long-range interactions drove the transition first-order.
In the present case, the transition is already first-order with
short-range interactions, so it seems rather likely that the
transition will remain so with long-range interactions.  This is
indeed what we find based on a renormalization group analysis.  We
will therefore only outline the calculation and state the results.

Consider Eq.\ (\ref{Phi2}) with $\Phi[n] = \int_{{\bf r}'} V({\bf r -
r'}) n({\bf r'})$ and $V(k) = 1/k$.  Once again, the $\lambda$
interaction is strongly relevant in two dimensions, so we would like
to continue Eq.\ (\ref{Phi2}) to $d$ dimensions.  We will only
consider the simplest isotropic continuation and assert that Eq.\
(\ref{Phi2}) holds in $d$ dimensions.  The upper critical dimension is
then $d_{uc} = 7$.  We have performed a one-loop $\epsilon$-expansion
in $d = 7-\epsilon$ dimensions, and find that the model lacks a stable
fixed point.  Thus, as expected, the transition remains first-order
when long-range interactions are included.

\section{Discussion and Summary}

The focus of this paper has been on the physics near the transition to
zero resistance state in 2DEGs driven with microwave radiation.  Our
goal was to understand the long-distance, long-time properties of the
system taking into account noise and fluctuation effects within a
nonequilibrium hydrodynamic theory involving the electron current and
density. We specifically focused on the transition to a
time-independent, density-ordered state that occurs when the
microscopic resistance first becomes negative at
$(\bk=0,\omega=0)$. The long wavelength subdiffusive density
fluctuations are the only critical modes at this transition. We
analyzed two models involving the density mode: (i) Model-I,
characterized by an imposed symmetry under a global uniform shift of
the density, valid only on sufficiently small length scales, and (ii)
Model-II, which is most general rotationally invariant model with no
additional symmetries.

The ordered state in Model-I consists of a uniform current and a
transverse Hall electric field that balances the Lorentz force.  This
state was shown to be stable within a linearized theory of
fluctuations about the ordered state.

We argued that the uniform-current steady state in Model-I cannot
exist in arbitrary large samples since the uniform Hall field would
eventually lead to regions of negative density.  Using parameters from
Willett's experiments\cite{Willett}, we estimated that samples with
dimension smaller $L_{\rm c1} \sim 1$mm can support this state.  To
describe the ordered state in larger samples, one must include terms
that depend on the magnitude of the density, which would prevent the
density from becoming arbitrarily large and negative.

Since the ordered state in Model-I breaks continuous rotational
symmetry, there is an associated Goldstone mode corresponding to
long-wavelength fluctuations of the current transverse to the uniform
current flow direction. We suggest that a possible way of detecting
this Goldstone mode might be to use surface acoustic waves in the zero
resistance regime.\cite{SAW1,SAW2} A surface acoustic wave at the
right wavelength and frequency should couple to this excitation,
leading to anomalous shifts in the velocity and intensity of the wave.

The transition to this ordered state in Model-I was analyzed both in
the case of short- and long-range interactions using dynamical
renormalization group methods. This model is valid for describing the
transition on length scales $L < L_{\rm c2}$, which we think could be
comparable to sample sizes in current experiments as discussed in
Section III~C, although it would be valuable to have an estimate from
microscopic calculations. With long-range interactions, we showed that
Model-I undergoes a first-order transition. However, with short-range
interactions, we showed that in two dimensions the Gaussian fixed
point in Model-I has a finite-volume basin of attraction. That is, a
finite-volume region of initial nonlinear couplings all flow to zero
upon renormalization. The transition in these cases is of the mean
field type. In particular, mean-field theory predicts a
\emph{continuous} transition to the ordered state.  Additionally, the
density subdiffuses at the critical point, with a frequency given by
$\omega
\propto -i k^4$.  This subdiffusion should lead to large density
fluctuations and hence large voltage fluctuations at the transition.
It may be interesting to observe this in samples with metallic gates,
so that the Coulomb interactions are screened.  Although it may be
difficult to quantitatively test the mean-field predictions, one could
perhaps measure voltage correlations at contacts placed along the
perimeter of the sample.  These voltages should behave similarly to
the density-density correlations given in Eq.\
(\ref{ncorr}). Qualitatively, one should at least observe large
voltage fluctuations since the density is critical and subdiffusive at
the transition.

We next turned to an analysis of the transition in the more generic
Model-II, which includes terms that depend explicitly on the magnitude
of the density.  We found that the transition within this model is
always first-order independent of whether interactions are short- or
long-range, at least near the upper-critical dimension of the theory.
The physical mechanism for this first-order transition is as follows.
If the resistance minimum occurs at $k = \omega = 0$, then one would
expect long-wavelength fluctuations to give to a time-independent
state with a uniform current and transverse Hall field.  As mentioned
above, such a state cannot exist in arbitrarily large samples since
regions of negative density would eventually appear.  The role of
terms that depend on the magnitude of the density is to prevent such
unphysical features from arising.  These terms consequently force a
first-order transition into a more complicated ordered state.

The experiments conducted so far were carried out using samples
without metallic gates, leading to unscreened Coulomb interactions.
The transition in these systems is therefore predicted to be
first-order, which should have measurable consequences.  In
particular, one would expect discontinuous jumps in various observable
quantities such as spontaneous currents, voltages, and local
magnetizations that develop at the transition.  We realize that these
jumps may be difficult to measure experimentally.  Another possibly
more controlled way of detecting a first-order signature might be to
measure the critical current above which the zero-resistance state
disappears.\cite{Andreev} If one approaches the transition from the
ordered state (by, say, changing the magnetic field) then the critical
current should drop discontinuously from some finite value to zero if
the transition is indeed first-order.

There are several future directions one could pursue with the theory
presented here that we believe would be interesting and provide
further insight into the remarkable physics of driven 2DEGs.
Regarding the transition to zero resistance, we have considered
only the simplest case where the resistance minimum occurs at $k =
\omega = 0$.  It may be interesting to generalize our results for this
case to include static disorder to see how it affects the transition.
One could also analyze the transition at finite frequency where a
time-dependent state such as circulating currents would arise.
Additionally, one could consider the transition at zero frequency but
nonzero wavevector $k_0$.  In this case one would be interested in
wavevectors $k$ such that $|k-k_0| < \Lambda$ for some cutoff
$\Lambda$.  If the equation of motion was derivable from a free energy
of the form
\begin{eqnarray}
  F &=& \int_{\bf q}r({\bf q})n({\bf q})n({\bf -q})
  \nonumber \\
  &+& \int_{{\bf q}_1,{\bf q}_2,{\bf q}_3} \lambda({\bf q}_1,{\bf
  q}_2,{\bf q}_3)
  n({\bf q}_1)n({\bf q}_2)n({\bf q}_3)
  \nonumber \\
  &+& \int_{{\bf q}_1\cdots{\bf q}_4} u({\bf q}_1,\ldots,{\bf q}_4)
  n({\bf q}_1)n({\bf q}_2)n({\bf q}_3)n({\bf q}_4)
  \label{eqF}
\end{eqnarray}
then we know that the cubic term drives the transition first-order
based on analogies with the solidification of an isotropic
liquid.\cite{Alexander} Even in the case where the cubic term
vanishes, the transition is still driven first-order by fluctuations
\cite{Brazovskii}.  Due to the presence of nonequilibrium terms,
however, the equation of motion will \emph{not} be derivable from a
free energy.  Nonequilibrium effects could cause dramatic deviations
from the equilibrium theory, and at present it is unclear what effect
such terms will have on the transition.

Another avenue one could pursue with this theory is to address the
properties of the ordered state away from the transition in Model-II.
Numerical studies may be best suited for this purpose especially since
the ordered state is likely to be inhomogeneous and not analytically
tractable.  One possible route is to generalize the numerics on the
flocking transition done by Vicsek \emph{et al}.\cite{Vicsek} to
include a magnetic field and interactions.

Finally, we note that while early numerical work on the flocking
transition in the absence of a magnetic field indicated a continuous
phase transition\cite{Vicsek}, some recent simulations on larger
system sizes hint at a weak first-order transition\cite{Gregoire}. If
true, this would be consistent with the transition continuing to be
first-order in the presence of a magnetic field for large enough
systems as argued in this paper.

\bigskip

\begin{acknowledgments}
The authors gratefully acknowledge Michael Cross, Jim Eisenstein, Matt
Foster, Hsiu-Hau Lin and R. Rajesh for useful discussions.  This work
was supported by the National Science Foundation through a Graduate
Research Fellowship (J.\ A.) and grants DMR-9985255 (L.\ B.\ and A.\
P.), PHY-9907949 (A.\ P.\ and M.\ P.\ A.\ F.), DMR-0210790 (M.\ P.\
A.\ F.), and DMR-0321848 (L.\ R.).  We also acknowledge funding from
the Packard Foundation (L.\ B., A.\ P., and L.\ R.) and the Alfred
P. Sloan Foundation (L.\ B.\ and A.\ P.).

\end{acknowledgments}


\begin{thebibliography}{30}
\bibitem[{not()}]{note}
\bibinfo{note}{The first order transition can be explicitly demonstrated for an
  equilibrium system, and is likely to remain so even in the nonequilibrium
  case.}
\end{thebibliography}

\end{document}